\documentclass[aps,pre,twocolumn,showpacs] {revtex4}
\usepackage{epsfig} \usepackage{bm} 
\begin{document}
\newcommand{\B}[1]{{\bm{#1}}}
\newcommand{\C}[1]{{\mathcal{#1}}} 
\renewcommand{\it}[1]{\textit{#1}}
\newcommand{\Onecol} {\begin{widetext} \onecolumngrid} 
\newcommand{\Twocol} {\end{widetext} \twocolumngrid}   
\title{Quasi-Gaussian Statistics of Hydrodynamic 
Turbulence in $\B {\frac{4}{3}+\epsilon}$  dimensions}

\author{Victor S. L'vov, Anna Pomyalov and Itamar Procaccia}   
\affiliation{~Department of Chemical Physics, The Weizmann Institute
of Science, Rehovot 76100, Israel} 
\begin{abstract}
The statistics of 2-dimensional turbulence exhibit a riddle: the
scaling exponents in the regime of inverse energy cascade agree with
the K41 theory of turbulence far from equilibrium, but the probability
distribution functions are close to Gaussian like in equilibrium. The
skewness $\C S \equiv S_3(R)/S^{3/2}_2(R)$ was measured as $\C
S_{\text{exp}}\approx 0.03$.  This contradiction is lifted by
understanding that 2-dimensional turbulence is not far from a
situation with equi-partition of enstrophy, which exist as true
thermodynamic equilibrium with K41 exponents in space dimension of
$d=\frac{4}{3}$. We evaluate theoretically the skewness $\C S(d)$ in
dimensions $\frac{4}{3}\le d\le 2$, show that $\C S(d)=0$ at
$d=\frac{4}{3}$, and that it remains as small as  $\C
S_{\text{exp}}$ in 2-dimensions.
\end{abstract}
\maketitle 

Two dimensional (2D) turbulence is not realized in nature or the
laboratory, but only in computer simulations \cite{80KM}. Nevertheless
it serves as an idealized model for a variety of natural flow
phenomena, like geophysical flows in the atmosphere, oceans and
magnetosphere. Experimental setups that are close to 2D turbulence
were realized in a number of laboratories \cite{Tabeling},
and the advent of faster computers allows precise simulations of the
2D Navier-Stokes equations \cite{00BCV}. In this Letter we are interested in
the statistical characteristics of 2D turbulence, in the probability
distribution functions and spectra of velocity differences.  The
velocity difference across a scale $\B R$ is written in terms of the
velocity field $\B u(\B r,t)$,
\begin{equation}
\B w(\B r,\B R,t)\equiv \B u(\B r+\B R,t)-\B u(\B r,t) \ . \label{diff}
\end{equation}
Usually one measures the longitudinal component,
\begin{equation}
w_\ell(\B r,\B R,t) \equiv [\B u(\B r+\B R,t)-\B u(\B r,t)]
\cdot {\B R}/{R} \ ,
\label{diff}
\end{equation}
the probability distribution function (pdf) of this object, denoted as
$P[w_\ell(\B r,\B R)]$, as well as moments of this pdf, like the
second and third order structure functions
\begin{equation}
S_2(R) \equiv \langle w^2_\ell(\B r,\B R,t)  \rangle \ , \quad
S_3(R) \equiv \langle w^3_\ell(\B r,\B R,t)  \rangle \ . \label{Sdef}\\
\end{equation}
Here the average is over space and time. In stationary homogeneous
and isotropic ensembles the structure function depends on $R$ only.

It is well known that in 2D turbulence the properties of these objects
depends on whether $R$ is larger or smaller than the scale $L$ at
which energy is injected into the system. For $R\gg L$ (but smaller
than the outer boundaries of the system) one observes an inverse
cascade of energy, and $S_2(R)$ is found to scale with an exponent in
agreement with the Kolmogorov 1941 theory, i.e.
\begin{equation}
S_2(R)=C_2 (\varepsilon R)^{2/3} \ , \quad R\gg L . \label{S2K41}
\end{equation}
Here $C_2$ is a dimensionless coefficient of the order of unity, and
$\varepsilon$ is the mean energy flux per unit time and mass.  For
$R\ll L$ (but larger than the dissipative scale) one observes a direct
cascade of enstrophy, with an exponent close to 2, but with some
logarithmic corrections. Recent experiments and simulations lend 
strong support to Eq. (\ref{S2K41}); on the face of it this indicates
that the system is very far from equilibrium, where equipartition of
energy is expected. On the other hand, the same experiments and
simulations indicate that $P[w_\ell(\B r,\B R)]$ appears almost
Gaussian, as if the system were very close to
equilibrium.  Quantitatively one measures the skewness
\begin{equation}
\C S = S_3(R)/S^{3/2}_2(R) \ , \label{skew}
\end{equation}
with the results that $\C S_{\text{exp}}\approx 0.03\ll 1$.  This
seeming contradiction and its resolution are the subjects of this
Letter.

The basic idea of this Letter is that there exists one space dimension
for which these observations are not at all in contradiction. This is
$d=\frac{4}{3}$, in which there exists an equilibrium state with
equipartition of enstrophy, where the scaling expected for $S_2(R)$ is
exactly (\ref{S2K41}), and where the skewness $\C S=0$.  By examining
turbulence in $\frac{4}{3}$+$\epsilon$ dimensions, we establish that
throughout the range $0\le \epsilon \le \frac{2}{3}$ the situation
remains very close to the one seen at $\epsilon=0$ ;  
the value of $\C S(d)$ provides a natural small parameter to
characterize the distance from $d=\frac{4}{3}$.  This parameter
remains very small  up to 2-dimensions. We thus interpret the
statistics in the inverse cascade regime of 2D turbulence as a state
very close to equilibrium.

It should be said at this point that attempts to connect turbulence to
equilibrium statistical mechanics were made before.  Well known are
the theories advanced by Onsager \cite{49On}, Hopf \cite{52Ho} and Lee
\cite{52Le}, and see ref. \cite{80KM} for a review. There ideas from
equilibrium statistical mechanics were proposed to explain turbulence;
as summarized in \cite{80KM}, the consensus is that such ideas were
not relevant for forced turbulence, but may be relevant for the final
stages of the temporal evolution of decaying turbulence.

Next there were attempts to connect the physics in 2-dimensions to
special properties of other, non physical dimensions. Of particular
influence were ideas advanced by Fournier and Frisch \cite{78FF}, who
identified $d\approx 2.05$ as the dimension at which the direction of
the energy flux changes sign. It was hoped that this may provide a
convenient starting point for perturbative expansions in the magnitude
of the flux. A similar point was identified in shell models of
turbulence, as a function of a parameter \cite{02GJY}. It turned out
however that the fluctuations were not small at that point, the flux
changed sign discontinuously, and perturbative theories did not yield
useful insights. In our own work on shell models \cite{02GLPP} we
identified another point in parameter space where a quasi-equilibrium
state with equi-partition of ``enstrophy" coincided with an energy
spectrum in agreement with Kolmogorov scaling exponents as in
(\ref{S2K41}). This has led us to seek an 1-parameter lifting of the
Navier-Stokes equations where a similar phenomenon exists. We were
thus led to examining turbulence in $d=\frac{4}{3}$ and its
vicinity. We argue below that there is a fundamental difference
between the properties of turbulence in $d=\frac{4}{3}$ (where the
flux vanishes) and $d\approx 2.05$ (where the flux discontinuously
changes sign). At the former point the statistics is exactly Gaussian,
with equipartition of the enstrophy.  The latter has no such
distinction, and the statistics may differ strongly from Gaussian.

We begin by generalizing 2D turbulence to $d<2$ dimensions. In doing so
we preserve the $d$-dimensional energy and enstrophy, in distinction
with other such extensions that fail to do so.  The Navier-Stokes
equations for the velocity field $\B u(\B r,t)$ read
\begin{equation}
 \label{NS1}
  \partial   {\B u}/ \partial t + (   {\B u}\cdot
  {\B {\nabla}})   {\B u} -\nu\nabla^2   {\B u}
 +{\B {\nabla }} p = {\B f}, \quad {\B {\nabla}}\cdot   {\B u}=0\,,
\end{equation}
where $\nu$ is the kinematic viscosity, $p$ is the pressure, and ${\bf
f}$ is a force which maintains the flow.  In $d=2$ it is
natural to consider the vorticity field
$\B \omega(\B r,t)=\B \nabla\times \B u(\B r,t)$.
For $\nu=\B f=0$ the curl of Eq. (\ref{NS1}) is
  $\partial   {\B \omega}/ \partial t= \B \nabla \times (\B u\times \B
  \omega)$.
Introduce the right-handed coordinate system ($x_1=x$, $x_2=y$ and
$x_3=z$) in which for 2D flows $u_3=0$ and $\B u(\B r,t)=\B
u(\B x,t)$, where $\B x=\{x_1,x_2\}$. The vorticity has the single
component $\omega_3$ for which Eq. (\ref{NS1}) reduces to
\begin{equation}
  \label{NS3}
  \partial {\omega}/ \partial t+ (\B u\cdot \B \nabla)\, \omega=0\ .
\end{equation}
This equation has two quadratic integral of motions, the (kinetic)
energy $E$ and the enstrophy $H$:
\begin{equation}
  \label{ints}
  E=\frac{1}{2}\int u^2(\B x,t)\, d^2 x\,, \quad H =\frac{1}{2} \int
  \omega^2(\B x,t)\, d^2 x\ .
\end{equation}
The velocity and vorticity of a 2D flow may be derived from the {\it
stream function} $\psi(\B x,t)$:
\begin{equation}
  \label{pot}
  \hskip -0.15cm \B u (\B x,t)=-\B \nabla \times \hat {\B z} \psi(\B
  x,t) \,, \quad \omega (\B x,t) = -\nabla^2 \psi(\B x,t)\,,
\end{equation}
where $\hat {\B z}$ is a unit vector orthogonal to the $\B x$ plane,
and $ \nabla^2$ is the Laplacian operator in the plane.  In $\B
k$-representation,
$
a(\B k,t)\equiv k\int d{\B r} \exp[-i({\B r}\cdot{\B k} )]
{\psi}(\B r,t)$ .
The Fourier transforms of $\B u(\B x,t)$ and of $\B \omega(\B x,t)$,
respectively $\B u(\B k,t)$ and $\B \omega(\B k,t)$, are expressed in
terms $a(\B k,t)$:
\begin{eqnarray}
  \label{u-omega}
  \B u(\B k,t)=i(\hat {\B z}\times \hat {\B k})\, a(\B k,t)\,,
  \quad \omega(\B k,t)= -k\, a(\B k,t)\,,
          \end{eqnarray}
where $\hat {\B k} =\B k/k$.  Now, by Eqs.~(\ref{NS3}) and
(\ref{u-omega})
\begin{eqnarray}
  \frac{\partial a(\B k,t) }{ \partial t}\!\!&=&\! \!\frac{1}{2}\int
  \!\!\frac {d^2 q d^2 p}{ (2\pi)^2} \delta (\B k+\B q+\B p)V_{kqp}~
  a^*(\B q,t)a^*(\B p,t),\nonumber \\ \label{NS4} 
V_{kqp}\!\!&=&\!\!  S_{kqp}(p^2-q^2)/2 kqp\,,\ S_{kqp} \equiv2 q p\,
\sin \phi_{pq}\,,
\\  \nonumber 
S_{kqp}\!\!&=&\!\!
S_{pkq}=S_{qpk}=- S_{kpq}=- S_{qkp}=- S_{pqk}\,,\\ \nonumber 
|S_{kpq}|\!\!&=&\!\!\sqrt{2(k^2q^2+q^2p^2+p^2k^2)-k^4-q^4-p^4}\
. \nonumber
\end{eqnarray}
Here the interaction amplitude (or ``vertex'') $V_{kqp}$ is expressed
via $S_{kqp}$; $|S_{kqp}|/4$ is the area of the triangle formed by the
vectors $\B k\,,\B q$ and $\B p$. $\phi_{pq}=\phi_p-\phi_q$; $\phi_k$,
$\phi_q$ and $\phi_p$ are the angles in the triangle plane between the
$x_1$-axis and the vectors $\B k$, $\B q$ and $\B p$ respectively.

The vertex $V_{kqp}$ satisfies two Jacoby identities~\cite{80KM}
\begin{eqnarray}
  \label{jac1}
  (V_{kqp}+V_{pkq}+V_{qpk})&=&0\,, \\
  (k^2V_{kqp}+p^2V_{pkq}+q^2V_{qpk})&=&0\ . \label{jac2}
\end{eqnarray}
The first one guarantees the conservation of energy in the inviscid
forceless limit, while the second conserves the
enstrophy. In terms of $a(\B k,t)$ Eqs. (\ref{ints}) read
\begin{equation}
  \label{ints2}
\hskip -0.3cm   E=\int\frac{|a(\B k,t)|^2 \, d^2 k}{2(2\pi)^2}\,, \ 
H=\int\frac{k^2\, |a(\B k,t)|^2\, d^2 k}{2(2\pi)^2}\ .
\end{equation}

For a statistical description in $d$-dimensions we consider the 2nd
and 3rd order correlation functions of $a,a^*$:
\begin{eqnarray}
  \label{corr2}
 \hskip -0.7cm (2\pi)^d \delta(\B k+\B q)n_k(t)&=&\langle a(\B k,t)
 a(\B q,t)\rangle\ , \\ 
 \hskip -0.7cm (2\pi)^d \delta(\B k\!+\!\B q\!+\!\B
 p)F_{kqp}(t)&=&\langle a(\B k,t) a(\B q,t) a(\B p,t)\rangle\
 . \label{corr3}
\end{eqnarray}
In isotropic systems, we do not need to carry the bold-face $\B k$
index in $n_k$ and in $F_{kqp}$.  In terms of $n_k$ we define the
volume densities in $d$ dimensions:
\begin{equation}
  \label{ints2}\hskip -0.1cm 
  \C E\equiv\frac{E}{V}=\int\frac{d^d k}{2(2\pi)^d}\,n_k\,, \ 
\C H\equiv \frac{H}{V}=\int\frac{d^d k}{2(2\pi)^d}\, k^2n_k\ .
\end{equation}
Since $n_k$ and $k^2 n_k$ serve as the energy and enstrophy densities
respectively in $\B k$-space, the thermodynamic equilibrium can be
achieved with equi-partition in any of these quantities:
\begin{eqnarray}\label{equipartition}
n_k^{\varepsilon\,0} &=&A_\varepsilon \ ,\qquad~~ \text{energy
equi-partition;}\\ \nonumber 
n_k^{h\,0} &=&A_h/k^2\ , \quad\text{enstrophy equi-partition} \ .
\end{eqnarray}
In such a state of thermodynamic equilibrium unavoidably all fluxes
vanish and the pdf of the velocity differences is Gaussian. Usually in
the theory of turbulence one rather considers flux equilibria, which
in the present situation can be with energy flux or enstrophy flux.
Dimensional analysis for such flux equilibria predicts
\begin{eqnarray} \label{flux-e}
\hskip -0.5cm n^{\varepsilon}_k=C_{\varepsilon}(d)
\varepsilon^{2/3}k^{-x_\varepsilon}\,,& \ x_\varepsilon=
d+\frac{2}{3}\,,&\ \text {energy flux;}\\ 
\hskip -0.5cm n^{h}_k= C_h(d)h^{2/3}k^{-x_h}\,, &\
x_h=d+2\,,&\ \text {enstrophy flux}. \label{flux-h}
\end{eqnarray}
Here $h$ is the mean enstrophy flux per unit time and mass, whereas
$C_\varepsilon(d)$ and $C_h(d)$ are $d$-dependent dimensionless
coefficients.  In terms of $S_2(R)$ these results are in agreement
with (\ref{S2K41}) for all $d$ for energy flux equilibrium. For
enstrophy flux equilibrium the result is $S_2(R)\propto R^2$.  The
basis for further development is the immediate observation that for
$d=\frac{4}{3}$ the scaling exponent for energy flux equilibrium,
$x_\varepsilon=2$, coincides with the the exponent 2 of the
equi-partition of enstrophy.  Accordingly for $d=\frac{4}{3}$ the law
(\ref{S2K41}) is in agreement with enstrophy equipartition, and
therefore also with a Gaussian pdf for the velocity differences.  In
the rest of this Letter we show that in dimensions $\frac{4}{3}<d\le
2$, the flux remains small (for given total energy of the system) and
the pdf's do not change much from Gaussianity.

A measure for deviations from Gaussianity is the skewness (\ref{skew})
which is now $d$-dependent, $\C S(d)$. To compute it we need to
separately find $S_2(R)$ and $S_3(R)$. We start with the former.  The
structure function $S_2(R)$ can be computed from Eqs. (\ref{u-omega})
and (\ref{flux-e}). In 2-dimensions
\begin{eqnarray}\nonumber 
 && \hskip -0.3cm 
S_2(R)=\int\frac{d^2k}{(2\pi)^2}|\exp(i k R \cos \phi_k )-1|^2
 \sin^2 \phi_k n^\varepsilon_k\\ \nonumber &=&\frac{C_\varepsilon 
\varepsilon
 ^{2/3}}{2\pi^2}\int\limits_0^{2\pi}\sin^2 \phi_k
 d\phi\int\limits_0^\infty \frac{d k} {k^{5/3}} [1-\exp(i k R \cos
 \phi_k )] \\ &=&C_\varepsilon (\varepsilon R) ^{2/3} A_2\,,
 \label{S2-nk}\\  \nonumber  
&& A_2=\int\limits_0^{2\pi}\frac{\sin^2 \phi_k
d\phi}{2\pi^2}\int\limits_0^\infty \frac{d \kappa } {\kappa^{5/3}}
[1-\cos(\kappa \cos \phi_k )]\\ \nonumber 
&=&\frac{ 27 \, \Gamma(2/3)}{ 2^{2/3}\, 32 \, \pi \,\Gamma(1/3)}\approx
0.0855\ .
\end{eqnarray}
In $d$ dimensions we write
\begin{equation}
S_2(R) =C_2(d)(\varepsilon R)^{2/3}
=C_\varepsilon(d) (\varepsilon R) ^{2/3} A_2(d)
\ .
\end{equation}
However, it is easy to see that $A_2(d)$ does not become
critical at $d=\frac{4}{3}$. On the other hand we will show
that $C_\varepsilon(d)$ (and therefore $S_2(R)$) diverges when $d\to
\frac{4}{3}$. Therefore we will estimate $A_2(d)$ by its 
value $A_2$ at 2-dimension, $C_2(d)=C_\varepsilon(d) A_2$.

On the one hand $S_3(R)=12\varepsilon R/d(d+2)$ exactly.
On the other hand the function $S_3(R)$ is
related to $F_{kqp}$ by: 
\begin{eqnarray}\nonumber 
&& S_3(R)=\int\frac{d^d k d^d q d^d p}{(2\pi)^{2d}} \delta(\B k+\B q+
\B p) \sin \phi_k \sin \phi_q \sin \phi_p\\ \nonumber 
&&\times \text{Im}\{ [\exp(i k R \cos
\phi_k )-1][\exp(i q R \cos \phi_q )-1]\\  
&&\times [\exp(i p R \cos \phi_p )-1]\}  F_{kqp} \ . \label{S3F3}
\end{eqnarray}
This is as far as we can proceed exactly. Now we will express the
third order correlator $F_{kqp}$ in terms of the second order $n_k$. It
is well known that this cannot be done without closure
approximations. The latter are known to provide semi-quantitative
estimates of the coefficients of correlation functions, and in the
present context we expect such approximations to perform better than
in 3D due to the existence of the small parameter $\C S(d)$ that we
will expose in this calculation. The starting point is the equation of
motion of $n_k(t)$, which can be exactly written in terms of the third
order correlation $F_{kqp}$:
\begin{eqnarray}
\partial n_k/2\,\partial t&=&I_k \ , \nonumber\\
I_k&=&\!\!\int
\!\frac {d^d q d^d p}{ (2\pi)^d} \delta (\B k+\B q+\B
p) V_{kqp} F_{kqp} \ . \label{gke}
\end{eqnarray}
A standard closure approximation expresses $F_{kqp}$ via
$n_k$. Proper closures in turbulence involve the
following steps: first, one considers the sweeping-free renormalized
perturbation theory to first order. Second, one assumes a
simple analytic form for the time-decay of correlation and response
functions.  A typical result reads \cite{71Kra,97LLNZ}
\begin{eqnarray}
  \label{closure}
F_{kqp}&=&N_{kqp}\theta_{kqp} \ , \\
N_{kqp}&=&V_{kqp}n_qn_p+V_{pkq}n_kn_q
+V_{qpk}n_pn_k\ . \nonumber
\end{eqnarray}
Here $\theta_{kqp}$ is a closure dependent ``triad-decorrelation
time''; assuming simple exponential decay for the 2nd order
correlation function and response function, one finds
\begin{equation}
  \theta_{kqp}= 1/(\gamma_k+\gamma_q+\gamma_p)\ .   \label{time}
\end{equation}
In Eq. (\ref{time}) $\gamma_k$ is the width of the assumed Lorentzian
line shape. The latter can be estimated as follows:
\begin{equation}
\gamma_k = A_\gamma(d) k \sqrt{S_2\left(1/k\right)}
=C_\gamma(d)\varepsilon^{1/3} k^{2/3} \ .
\label{gammak}
\end{equation}
Here $A_\gamma(d)$ is a coefficient of the order of unity. Since
$S_2(R)$ diverges at $d=\frac{4}{3}$ we will evaluate
below $A_\gamma(d)$ via its 2D value $A_\gamma$:
\begin{equation}
C_\gamma(d)=A_\gamma\sqrt{C_2(d)}\ . \label{CgamAgam}
\end{equation}

We should stress that although Eq. (\ref{closure}) is derived using
usual uncontrolled closure approximations, once it is substituted into
Eq. (\ref{gke}) the latter conserves the energy and enstrophy
invariants defined in Eq.(\ref{ints2}) in all $d$ dimension. This
distinguishes our analysis from some previous theories like
\cite{78FF} which conserved enstrophy in 2-dimensions only. Note that
our equations of motion exhibit the equilibria (\ref{equipartition})
and (\ref{flux-e}), (\ref{flux-h}) as exact results. Thermodynamic
equilibria follow directly from the Jacoby identities (\ref{jac1})
and (\ref{jac2}) which yield  
$N_{kqp}=F_{kqp}=0$ and hence also $I_k=0$.  To show that also the flux
equilibria are satisfied exactly one needs to use the
Kraichnan-Zakharov transformation \cite{ZLF} and the  Jacoby
identities. There $F_{kqp}$ does not vanish in general.

We now note that in dimension $d=\frac{4}{3}$ when $n_k\propto 1/k^2$,
$F_{kqp}$ vanishes by itself. This follows from the fact that
$N_{kqp}$ vanishes due to the Jacoby identity (\ref{jac2}). It is
worthwhile to stress that in our context this result is derived only
to first order in renormalized perturbation theory. It is however a
stronger result which can be established order by order to all orders.

Substituting (\ref{closure}) with $n_k=n_k^\varepsilon$ of 
Eq. (\ref{flux-e}), and Eq. (\ref{time}), we
can rewrite $S_3(R)$ in the form
\begin{equation}
S_3(R)=[C_\varepsilon(d)]^2 A_3(d)
(\varepsilon R) /C_\gamma(d)\ , \label{S3d}
\end{equation}
where $A_3(d)$ is:
\Onecol
\begin{eqnarray}\nonumber
A_3(d)&=&\int\frac{d^d \tilde k d^d\tilde q d^d\tilde p}{ (2\pi)^{2d}}
\frac{\delta(\tilde{\B k} +\tilde{\B q}+ \tilde{\B p}) S_{\tilde
k\tilde q\tilde p }}{\tilde k^{2/3} +\tilde q^{2/3}+\tilde p^{2/3}}
\frac{\sin \phi_k \sin \phi_q \sin \phi_p} {(\tilde k \tilde q \tilde
p)^{2d+5/3}} 
\big[ \sin (\tilde k \cos \phi_k )+\sin( \tilde q \cos \phi_q ) +\sin
(\tilde p \cos \phi_p )\big]
 \\   \label{A3d} 
&&\times   2\, 
[\tilde k^{d+2/3}(\tilde p^2-\tilde q^2)
+\tilde p^{d+2/3}(\tilde q^2-\tilde k^2)
+ \tilde q^{d+2/3}(\tilde k^2-\tilde p^2)]\ .
\end{eqnarray}
\Twocol
\noindent
Combining Eqs. (\ref{skew}),
(\ref{S2-nk}), (\ref{CgamAgam}) and (\ref{S3d}) we find
\begin{equation}
  \C S(d) =A_3(d)/A_2^2 A_\gamma 
\end{equation} 
Obviously, $A_3(d)=0$ at $d=\frac{4}{3 }$. This implies that $S_2(R)$
(i.e. $C_\varepsilon(d)$) diverge at this point.  It is easy to prove
that the first derivative of $A_3(d)$ with respect to $d$ at
$d=\frac{4}{3}$ is finite. We can therefore approximate $A_3(d)$ as
\begin{equation}
A_3(d) \approx a (d-4/3) \ , \label{A3dresult}
\end{equation}
up to orders of $(d-\frac{4}{3})^2$. On the other hand  by direct
numerical integration we found
\begin{equation}
A_3\equiv A_3(d)\big|_{d=2}\approx 3.556 \times 10^{-4} \ . \label{A3}
\end{equation}
Thus in 2-dimensions we estimate
\begin{equation}
\C S \equiv \C S(d)\big|_{d=2}\approx 0.0486 /A_\gamma \ .
\end{equation}
The experimental observation is that $\C S_{\text{exp}} \approx
0.03$. Taking liberty to use this finding we estimate $A_\gamma\approx
1.62$. This is in agreement with our expectation that $A_\gamma$ is of
the order of unity. Now we can estimate $\C S(d)$ in the whole range
$\frac{4}{3}\le d\le 2$ by using the linear approximation
(\ref{A3dresult}), and find
\begin{equation}
\C S(d)\approx
  \frac{0.0729}{A_\gamma} (d-\frac{4}{3})\approx 0.045 (d-\frac{4}{3}) .
\end{equation}
The main conclusion of this Letter is that although $d=2$ is finitely
removed from $d=\frac{4}{3}$, the relevant small parameter remains
small all the way to $d=2$, because of the numerical smallness of the
ratio $A_3/A_2^2\approx 0.0486$. This smallness stems from generic
geometric cancellations in the last line of the integrand in
(\ref{A3d}).  This originates from the structure of the vertex
$V_{kqp}$ and is therefore fundamental to Euler equations in
2-dimensions.

In summary, we have addressed the experimental findings of the
statistics of 2D turbulence. It is hardly surprising that 2D
turbulence is not strongly intermittent; in the inverse cascade regime
there are no mechanisms to create rare events that are intimately
related to the sharpening and enhancement of fluctuations as they are
transferred {\em down} the scales in generic 3D direct cascades
\cite{01LPP}. On the other hand the fact that the statistic of 2D
turbulence are so close to Gaussian came as a major surprise. In this
Letter we offered an explanation to this finding. We have
identified $d=\frac{4}{3}$ as a convenient point around which to develop a
theory of 2D turbulence. The statistics there are Gaussian, but the
spectrum in the inverse energy flux regime is K41. The skewness is zero,
allowing a sensible closure theory for
$d$ slightly larger than $\frac{4}{3}$. Estimating the skewness as a
function of $d$, we are led to conclude that it remains small also in
2-dimensions. We can thus interpret 2D turbulence as a state close to
equilibrium. In future work we will examine further the structure of
the theory in the range $\frac{4}{3}\le d\le 2$ to assess further the
quality of the closure approximation.

We thank R. H. Kraichnan for useful correspondence. This work was supported
by the Israel Science Foundation, the European Commission under a TMR
grant and the
Naftali and Anna Backenroth-Bronicki Fund for Research in
Chaos and Complexity.

\end{document}